\documentclass[prb,twocolumn,amsmath,amssymb,showpacs]{revtex4}
\usepackage{graphicx}% Include figure files
\begin{document}
%\preprint{APS/123-QED}

\title{Comment on the "Origin of the opalescence at the
$\boldsymbol{\alpha\leftrightarrow\beta}$ transition of quartz:
role of the incommensurate phase studied by synchrotron
radiation"}
\author{T.A. Aslanyan}
\email{aslanyan@freenet.am} \affiliation{Institute for Physical
Research, Armenian National Academy of Sciences, Ashtarak-2,
378410 Armenia}

\author{T. Shigenari}%
\author{K. Abe}
\author{D.A. Semagin}
\affiliation{Dept. of Appl. Phys. and Chem., University of
Electro-communications, Chofu-shi, Tokyo 182, Japan}

%\affiliation{

%\date{ }
\received{}
\begin{abstract}
The interpretation of the origin of the light scattering anomalies
near the transition from incommensurate (IC) to the $\alpha$-phase
of quartz by Dolino {\it et al.} (Phys. Rev. Lett. 94, 155701 (2005)) is
commented. It is shown that the observed IC structure is a pure
transverse acoustic (TA) modulation without any soft optic mode
component. Such a modulation cannot be responsible for the
observed light scattering contrary to the interpretation by Dolino {\it et al}.
\end{abstract}
%Use showkeys class option if keyword
\maketitle The present paper is a comment on the interpretation by
Dolino {\it et al.} \cite{1} of their recent observations in
quartz. In the paper\cite{1}, from the diffraction experiment
using synchrotron radiation, the authors showed that the
incommensurate (IC) modulation wavevector in quartz becomes
extremely small in the region of the sample close to the boundary
between the IC phase and the $\alpha$- phase. In this region,
where the two phases coexist in the same sample, very intense
light scattering is observed.\cite{shustin} Therefore, it is very
important to consider how the intense light scattering is related
to the reported extraordinary long period modulation.

In the following, we show that (i) the observed  long-period IC
modulation in quartz is a pure transversal acoustic (TA)
modulation, including neither longitudinal acoustic (LA) component
nor the optic mode component. (ii) A long-period acoustic
modulation cannot appear spontaneously as a result of the IC phase
transition associated with the acoustic mode softening, and the
mechanism of its formation should be understood. (iii) Such a
modulation cannot be at the origin of any observed light
scattering anomalies contrary to the interpretation by Dolino {\it
et al.}\cite{1}.

\begin{figure*}
\rotatebox{0}{\includegraphics* [scale =0.7]{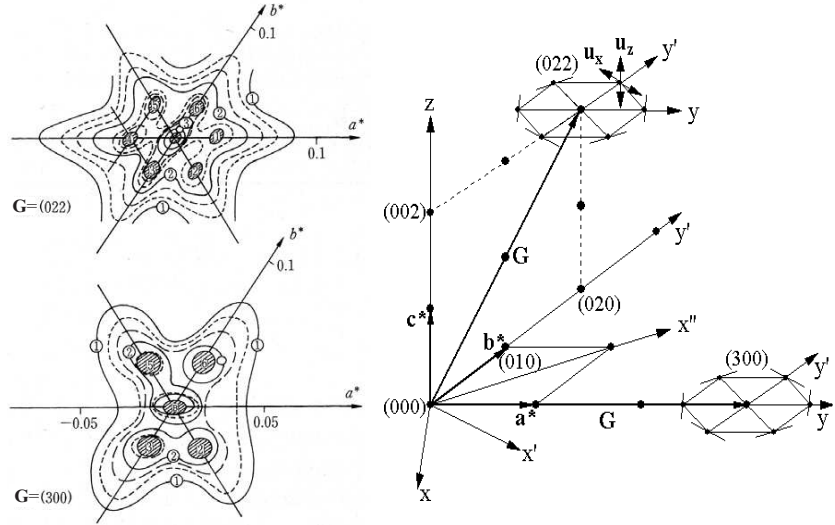}} \caption{IC
satellite reflections observed near the main Bragg reflections
${\mathbf G}$=(022) and (300) by Dolino et al\cite{3}. The two
missing satellites along the vector ${\mathbf G}$=(300), which is
parallel to the IC vector ${\bf k}$ (i.e. along the $a^{\ast}$
axis), make an evidence that the acoustic displacements in the IC
wave are perpendicular to the ${\mathbf G}$-vector, and the optic
displacements along with the LA displacements are zero in the IC
structure. As shown in the present paper, the missing satellites
should be observed as the most intensive among the six satellites
near the transition to the $\alpha$-phase, if the interpretation
by Dolino {\it et al.}\cite{1} is acceptable. The satellites along
the $a^{\ast}$ axis near (022)-reflection are the contribution
from the $z$-component of IC displacements wave of the TA type,
since the contribution of the optical displacements to these
positions is zero \cite{7}. On the right part, the acoustic
displacements with respect to the crystal's coordinate axes are
depicted.}
\end{figure*}

The consideration below is mainly based on the observations by
Dolino \textit{et al.} \cite{3} of the IC satellite reflections in
the elastic neutron diffraction, and their expected evolving in
view of the recent observations in the synchrotron radiation
\cite{1}. As observed (see Fig. 1), near the Bragg reflections
(100), (200) \ldots  the satellites, corresponding to the IC
vector ${\bf k}$ parallel to the main Bragg reflection vector
(l00), are systematically absent, i.e. four satellites are
observed instead of six. Similar observations by Gouhara and Kato
\cite{5,6} brought them to a conclusion, that the IC modulation in
quartz is mainly of an acoustic character, with a small (of about
0.1 fraction) optical atomic displacements component. The analysis
carried out by the present authors \cite{7}, based on the
comparison of the IC satellites near the (l00) and (l0m)-type
Bragg reflections, showed that the IC modulation in quartz is a
pure TA modulation, including, as one of its components, the
transversal $u_z$-displacements. The contribution of the TA
$u_z$-displacements to the diffraction was interpreted by Gouhara
and Kato \cite{5,6} as that from a small optical component. The
existence of $u_z$-displacements was also demonstrated by the
recent MD calculations. \cite{Dmitirev} Below we bring new
arguments, unambiguously demonstrating the TA character of the IC
modulation in quartz. However, contribution of a long-period
acoustic modulation to the dielectric constant's components (i.e.
to the light scattering) is proportional to the square of the IC
wavevector, and therefore it must be negligibly small for the
small IC modulation vectors observed in quartz ($k<0.03a^\ast$ and
decreases on cooling, according to the recent
observations\cite{1}, down to $0.002a^\ast$).

We show that in the case if the IC modulation contains an optic
mode displacements component, an intense IC satellite reflections
should necessarily appear in the positions of the missing
satellites (in paricular, near (300) in Fig. 1).

Let us assume that the IC modulation in quartz, however, contains
an optical component. In such a case, the long-period triple-$k$
IC modulation of the quartz's $\alpha\leftrightarrow\beta$
transition parameter $\eta$ necessarily should induce a
longitudinal acoustic (LA) modulation with a large amplitude
$u_{l0}$. The amplitude $u_{l0}$ of such LA modulation should be
of the same order of magnitude as that for the $\eta$- modulation
for the IC vectors $k\approx 0.03a^\ast$, and for the IC vectors
$k\approx 0.002a^\ast$ (near the transition to the $\alpha$-phase)
it should increase up to $\sim 10$ times (the corresponding
diffraction satellites should grow in intensity up to $\sim 10^2$
times). As a result, six intense IC satellites should be induced
by the LA modulation instead of the four satellites near any (l00)
reflection.

The elastic strain dependent part of the quartz's thermodynamic
potential is of the form \cite{4}:
\begin{eqnarray}
\nonumber \int\{ a\left[({u_{xx} }-{ u_{yy} })
\frac{\partial\eta}{\partial x}-{2u_{xy}
}\frac{\partial\eta}{\partial y}\right]+ r\eta^2(u_{xx}+u_{yy}) +
\\ \frac{c_{11}\! - c_{66}}{2}(u_{xx}\!+\!u_{yy})^2 + \frac{c_{66}}{2}\left
[ (u_{xx}\!-\!u_{yy})^2 + 4u_{xy}^2\right ]\}dV
\end{eqnarray}
where $u_{xy},~ u_{xx},~ u_{yy}$ are the elastic strains, $a$ and
$r$ are some coefficients of expansion, which are of an atomic
order of magnitude. For the triple-$k$ IC optical modulation
\[
\eta (R)=\eta_0\cos ({\bf kR}+\varphi )+\eta_0\cos ({\bf k_1R}+\varphi_1 )+
\eta_0\cos ({\bf k_2R}+\varphi_2 )
\]
with ${\bf k}+{\bf k}_1+{\bf k}_2=0$, and for the acoustic displacements vector
${\bf u}={\bf u}_l+{\bf u}_t$, presented as the sum of the longitudinal ${\bf u}_l$ (along
the IC vector $k$) and transversal ${\bf u}_t$ displacements, one can
minimize Eq. (1) with respect to ${\bf u}_l$, and obtain an equilibrium
LA modulation wave:
$${\bf u}_l={\bf u}_{l0}\sin ({\bf kR}+\varphi_1+ \varphi_2)+{\bf u}^\prime_{l0}
\cos ({\bf kR}+\varphi),$$
with the amplitudes $$u_{l0}=
\frac{2r\eta_0^2}{c_{11}k} \;\;{\rm and}\;\;u^\prime_{l0}=
\frac{a\eta_0 \cos 3\phi}{c_{11}},$$ where $\phi$ is the angle
between the $x$ axis and the IC vector ${\bf k}$. In the observed IC
structure the angle $\phi$ is close to $\pi /6$, and therefore the
corresponding amplitude $u^\prime_{l0}$ is close to zero and
generally neglected. An identical  LA IC amplitudes must exist for
all the six wavevectors $\pm {\bf k},\; \pm {\bf k}_1$ and $\pm {\bf k}_2$.

Taking into account that the IC wavevector in quartz is very
small, and it decreases on cooling from 0.03$a^{\ast}$ (as in
Fig. 1) down to 0.002$a^{\ast}$ (according to the recent
publication by Dolino\cite{1}), one can see the enormous increase
of LA modulation's amplitude $u_{l0}$ (see the $k$ vector in the
denominator).

The physical reason of the large LA amplitude is very plain.
Formation of a long-period acoustic modulation requires a very
small energy (this energy is proportional to $k^2$, and is tending
to zero with $k\to 0$). Though the acoustic-optic coupling term
$\eta^2u_{ii}$ in Eq. (1) is rather small, however it is sufficient
for inducing an enormous LA amplitude for the small $k$-s.

For calculation of the X-ray diffraction scattering amplitude,
induced by the IC wave, one should expand the crystals density
function as:
\begin{eqnarray}
\nonumber &&F_G\exp[i\mathbf{G}(\mathbf{R+u})]\approx
F_G\exp[i\mathbf{G} \mathbf{R}]+\\
&&iF_G(\mathbf{Gu}_0)\exp[i(\mathbf{G\pm k})\mathbf{R}],
\end{eqnarray}
where $F_G$ is the structure factor, corresponding to the $\mathbf
G$-Bragg reflection, ${\bf u}_0$ is the amplitude of the IC acoustic
modulation ${\bf u}({\bf R})$, and the scalar product
$(\mathbf{Gu}_0)=u_{l0}G$ for the case, when vector ${\bf G}$ is
parallel to the IC vector ${\bf k}$ (as for the case of the missing
satellites in Fig. 1). The first term in Eq. (2) gives diffraction
to the main Bragg reflection $\mathbf G$ with intensity $\sim
|F_G|^2$. The second term gives the acoustic modulation's
contribution to the satellite reflections $\mathbf{G\pm k}$ with
intensity $\sim |F_G(\mathbf{Gu}_0)|^2$. For the case of $\mathbf
G\parallel {\bf k}$, this intensity takes the form $|F_G Gu_{l0}|^2$,
and it extremely increases with $k\to 0$, since, as shown above,
$u_{l0}$ increases as $1/k$.

For the case $k=0.03a^*$, it is easy to estimate that the
satellite intensity  induced by $u_{l0}$ and contributing to the
position of the missing satellites in Fig. 1 should be of the same
order of magnitude as any intensity, induced by the optical
$\eta$-atomic displacements. Note that the satellite intensity
contributed from the optical component is $\sim |F_G G\eta_0|^2$.
In the vicinity of $1K$ of the phase transition, the optical
amplitude $\eta_0\sim 10^{-2}\eta_{at}$, where $\eta_{at}$ is of
an atomic order of magnitude, $k\sim 10^{-2}a^*$, and
subsequently, $u_{l0}$ and $\eta_0$ are of the same order of
magnitude. For the smaller IC vectors $k\approx 0.002a^*$,
$u_{l0}$ is larger than $\eta_0$ about $\sim 10$ times, and the
ratio of the corresponding intensities should make $\sim 10^2$. In
other words, for $k\leq 0.03a^*$ six reflections of about the same
intensity in Fig. 1 should be observed. And besides, since the
scalar product $(\mathbf{Gu}_0)$ in Eq. (2) is maximal for the
case when ${\bf G}$ and ${\bf k}$ are parallel, the intensity of
the $\mathbf{G\pm k}$ satellites should be larger, than the
intensity of the $\mathbf{G\pm k}_1$ and $\mathbf{G\pm k}_2$
satellites. In other words, the missing satellites in Fig. 1
should be detected as the most intensive, compared with four other
satellites near (l00).

The only explanation of the systematic observation of four
satellites near the (l00) type Bragg reflections is the TA
character of the IC modulation, without optical and LA components,
since otherwise, any optical component in the triple-k modulation
necessarily induces a strong LA component, which will contribute
to the positions of the missing satellites.

The TA IC structure, which, in fact, exists in quartz, cannot
noticeably contribute to the light scattering anomalies, since its
effect on the dielectric constant $\varepsilon_{ij}$ is
proportional to $u^2_{xy}$ (i.e. to a very small parameter $k^2$),
while the light scattering anomalies are observed just when $k\to
0$. So, we do not agree with the interpretation of the light
scattering anomalies in quartz by Dolino {\it et al.} \cite{1},
attributed to the observed IC modulation. Neither the domains of
the IC structure (the adjacent domains with different values of
$k$ in the small-angle scattering zone ~\cite{1}), nor the
Dauphine twins observed in the "fog zone" can be responsible for
the light scattering in quartz. The latter was discussed in detail
and estimated by the present authors earlier \cite{8}, where the
origin of the intense light scattering was attributed to the
ferroelastic domains. The arguments given above completely
contradict the observations and reject the traditional model for
the IC transition adopted by Dolino {\it et al.} \cite{1}.

Finally we introduce another argument, supporting the above
discussion. As it follows from the symmetry, in case of the
triple-$k$ IC optical modulation, the dielectric constant's
components $\Delta\varepsilon_{xx},\;\Delta\varepsilon_{yy},\;
\Delta\varepsilon_{zz}$ are proportional to $\sim\eta_0^2\sin
({\bf kR}+\varphi_1+ \varphi_2)$. Since the magnitude of the IC
vector ${\bf k}$ decreases down to 0.002$a^\ast$, then the light
of 500nm or smaller wavelength (i.e. with wavevector $\geq
0.001a^\ast$), should diffract in back-direction (or close to it)
in such IC modulation. Such a feature, which can be visually
detected, is not observed in quartz. This fact also proves, that
the LA modulation (and subsequently an optical modulation), which
directly follows from the traditional model used by Dolino
\textit{et al.} \cite{1}, does not exist in the IC phase of
quartz.

It should be mentioned that a long-period acoustic modulation
cannot appear spontaneously, as a result of the IC phase
transition, associated with the acoustic mode softening. Such a
mode softening would have a lot of consequences, including a
significant drop in the elastic constant $c_{66}$, which have
never been observed in quartz.\cite{kimizuka} One may assume that
some latent (driving) phase transition takes place in quartz near
the $\alpha\leftrightarrow\beta$ transition, and the observed
acoustic modulation is only a manifestation of this hidden
transition. However, a pure TA character of the IC modulation in
quartz was sufficiently simply and reliably demonstrated also in
the earlier consideration \cite{7}, based on the different
arguments. So, it cannot be ignored, and understanding of its
origin is of a priority importance in the quartz's phase
transition problem.

\end{document}